\documentclass[%
 reprint,
 amsmath,amssymb,amsfonts
 aps,
prb,
]{revtex4-2}
\usepackage[dvipdfmx]{graphicx}
\usepackage{graphicx}% Include figure files
\usepackage{dcolumn}% Align table columns on decimal point
\usepackage{bm}% bold math
\usepackage{txfonts}
\usepackage{color}
\usepackage{color}
\usepackage{here}
\usepackage{dcolumn}% Align table columns on decimal point
\usepackage{bm}% bold math
\usepackage[version=3]{mhchem}
\usepackage{mathtools}
\usepackage{newtxtext}
\usepackage[varg]{newtxmath}
\usepackage[normalem]{ulem}
\usepackage{lipsum}
\usepackage{braket}
\usepackage{comment}
\usepackage[final]{hyperref}
\usepackage[mathlines]{lineno}% Enable numbering of text and display math
\begin{document}

%\preprint{APS/123-QED}

\title{Finite-temperature second-order perturbation analysis { of} magnetocrystalline anisotropy energy of $\textrm{$L$1}_{0}$-type ordered alloys }% Force line breaks with \\

\author{Shogo Yamashita}\email{shogo.yamashita.q1@dc.tohoku.ac.jp} 
 \author {Akimasa Sakuma}

\affiliation{Department of Applied Physics, Tohoku University, Sendai 980-8579, Japan}%


%\affiliation{
%\affiliation{
 %Third institution, the second for Charlie Author
%\affiliation{%


\date{\today}% It is always \today, today,

\begin{abstract} We present a novel finite-temperature second-order perturbation method incorporating spin-orbit coupling to investigate the temperature-dependent site-resolved contributions to the magnetocrystalline anisotropy energy (MAE), specifically $K_{1}(T)$, in FePt, MnAl, and FeNi alloys. Our developed method successfully reproduces the results obtained using the force theorem from our previous work. By employing this method, we identify the key sites responsible for the distinctive behaviors of MAE in these alloys, shedding light on the inadequacy of the spin model in capturing the temperature dependence of MAE in itinerant magnets. Moreover, we explore the lattice expansion effect on the temperature dependence of on-site contributions to $K_{1}(T)$ in FeNi. Our results not only provide insights into the limitations of the spin model in explaining the temperature dependence of MAE in itinerant ferromagnets but also highlight the need for further investigations. These findings contribute to a deeper understanding of the complex nature of MAE in itinerant magnetic systems.
%\begin{description}
%\end{description}
\end{abstract}

\maketitle

%%%%%%%%%%%%%%%%%%%%%%%%%%%%%%%%%%%%%%%%%%%%%%%%%%%%%%%%%%%%%%%%%%%%%%%%%%
%%%% イントロ
%%%%%%%%%%%%%%%%%%%%%%%%%%%%%%%%%%%%%%%%%%%%%%%%%%%%%%%%%%%%%%%%%%%%%%%%%%
\section{Introduction}
The magnetocrystalline anisotropy energy (MAE) is { an} important characters of magnetic materials because it {governs} the coercivity. Rare-earth permanent magnets are examples of magnets { with}  the high coercivity and {are} used for many { modern applications}. 
{Recently, however,} the development of rare-earth{-}free magnets has been accelerating to avoid { the use of }expensive rare-earth elements. $L1_{0}$-type transition metal alloys such as FeNi are examples of rare-earth-free high-performance permanent magnets\cite{SRFeNi}. Generally, coercivity has {strong} temperature dependence{; thus,} we need to understand the temperature dependence of the MAE for { the} development of high-performance transition-metal magnets at finite temperatures. \\
\ The MAE for the uniaxial crystal $E_{\textrm{MAE}}(T,\theta)$ is usually expressed as follows:
\begin{align}
E_{\textrm{MAE}}(T,\theta)=K_{1}(T) \textrm{sin}^2\theta +K_{2}(T) \textrm{sin}^4\theta+\cdots,
\label{MAE}
\end{align}
where $K_{1}(T)$ and $K_{2}(T)$ are the anisotropy constants.
However, {the theoretical description of the temperature dependence of the MAE remains} controversial. In localized electron systems, for instance, $4f$ electron systems { such as} permanent magnets,  theories based on the localized spin model combined with the crystal field theory have been well-established
\cite{Herbst,Yamada,Sasaki2015,Yoshioka2018,Yoshioka2020,Yamashita2020,Yoshioka2022}. The Callen--Callen power law\cite{Akulov,Zener,Callen1,Callen2} is {a line} of these theories { that} can describe the temperature dependence of $K_{1}(T)$ and $K_{2}(T)$. In contrast, the cases of transition-metal magnets, which are treated as itinerant electron systems, are debatable. At 0 K, the mechanism of the MAE in the itinerant electrons systems, { particularly} $K_{1}(0)$, can be explained by the second-order perturbation formula {in terms of the spin -orbit coupling (SOC)} { according to} the tight-binding model\cite{Bruno,Laan,Kota1,Solovyev,Ke1,Ke2}.  
However,  finite-temperature expressions for the MAE based on the band theory { are} not available yet. This is because the band theory is based on the mean field theory and cannot describe the spin-transverse fluctuations directly. One of the ways to describe the spin fluctuation based on the itinerant electron theory is the functional integral method\cite{Cyrot,Hubbard1,Hubbard2,Hasegawa1,Hasegawa2,Hasegawa3}. This method is usually combined with coherent potential approximation (CPA).
In this  { approach}, the spin fluctuation can be expressed {by} random spin states with respect to its direction, which are called disordered local moment (DLM) states.
 { First-principles} calculations based on this scheme have been performed by several authors\cite{Oguchi,Pindor,Staunton3,Gyorrfy,StauntonPRL1992}  to investigate the finite-temperature magnetic properties of magnetic materials as pioneering works. Subsequently, the temperature dependences of the MAE, transport properties, and Gilbert damping constants in the itinerant electron systems  { were} investigated  { via the} DLM-CPA method  { and} the density functional theory\cite{StauntonPRL,StauntonPRB,Deak1,Matsumoto,Hiramatsu1,Sakuma2022,Yamashita2022,Juba1,Hiramatsu2023},  { along with} model calculations\cite{Sakuma2018,Miura2021,Miura2022}. 
 { In particular}, the temperature dependence of the MAE for $L1_{0}$-type alloys has been calculated by several authors\cite{StauntonPRL,StauntonPRB,Deak1,Yamashita2022}. \\ 
\ Recently, {we} calculated the temperature dependence of the MAE for $L1_{0}$-type FePt, MnAl, and FeNi using the DLM-CPA method{\cite{Yamashita2022}}.
 { The calculation} results for FePt and MnAl  { indicated} that the MAE decreases with {an increase in the temperature}. However, the calculated MAE for FeNi {exhibited a unique behavior}. It { did} not decrease monotonically with { an increase in the temperature; rather, it exhibited plateau-like behavior in the} low-temperature region. This behavior is similar to that of $\textrm{Y}_{2}\textrm{Fe}_{14}\textrm{B}$\cite{Juba1,Sagawa,Grossinger}, for which the mechanism of the temperature dependence of the MAE has been controversial. 
\\ \ {In the present study, to analyze these behaviors, we decompose the MAE of these alloys at finite temperatures into onsite and pair contributions { using} the second-order perturbation (SOP) method in terms of the SOC.}
In this method, we can extend the formula to describe $K_{1}(0)$ in the tight-binding model\cite{Bruno,Laan,Kota1,Solovyev,Ke1,Ke2} to the finite-temperature expression $K_{1}(T)$. 
%%%%%%%%%%%%%%%%%%%%%%%%%%%%%%%%%%%%%%%%%%%%%%%%%%%%%%%%%%%%%%%%%%%%%%%%%%
%%%% 式の説明
%%%%%%%%%%%%%%%%%%%%%%%%%%%%%%%%%%%%%%%%%%%%%%%%%%%%%%%%%%%%%%%%%%%%%%%%%%
\section{Calculation details}
 { To develop the finite-temperature {SOP} formula, we use the  tight-binding linearized muffin-tin orbital (TB-LMTO) method\cite{Andersen,Skriver,Turek,Kudrnovsky,Sakuma2000} with the atomic sphere approximations combined with the DLM-CPA method.} First, in the DLM-CPA method, we need to calculate a distribution function $\omega(\lbrace{\bf{e}}\rbrace,T)$ representing the probability with which the spin vectors are directed to $\lbrace{\bf {e}}\rbrace$ at temperature $T$.
In this work, we adopt the single-site approximation; thus, $\omega(\lbrace{\bf{e}}\rbrace,T)$ is decoupled { into the} simple product of the probability at each site $\omega_{i}({\bf{e}}_{i},T)$ as follows:
\begin{align}
\omega(\lbrace{\bf{e}}\rbrace,T)=\prod_{i} \omega_{i}({\bf{e}}_{i},T).
\end{align} 
In a previous work, $\omega_{i}({\bf{e}}_{i},T)$ { was} evaluated { using} the analogy of the Weiss field\cite{Gyorrfy,StauntonPRL,StauntonPRB,Deak1}. However, in this work, we determine $\omega_{i}({\bf{e}}_{i},T)$ by evaluating the effective grand potential $\Omega_{\textrm{eff}}(\lbrace{\bf{e}}\rbrace,T)$ of electrons. Here, we explain the { calculation of} $\Omega_{\textrm{eff}}(\lbrace{\bf{e}}\rbrace,T)$ and $\omega_{i}({\bf{e}}_{i},T)$.
First, we introduce the Green function including the spin-transverse fluctuation at finite temperatures in the TB-LMTO method $G(z, \lbrace {\bf{ e}}\rbrace)$\cite{Hiramatsu1,Sakuma2022,Yamashita2022,Hiramatsu2023}  as follows:
\begin{align}
G_{ij}(z;\lbrace {\bf{ e}}\rbrace)=\lambda^{\beta}_{i}(z; {\bf{ e}}_{i}) \delta_{ij}+ \mu^{\beta}_{i} (z;  {\bf{ e}}_{i}) g^{\beta}_{ij}(z;\lbrace {\bf{ e}}\rbrace) \bar \mu^{\beta}_{j} (z;  {\bf{ e}}_{j}),
\end{align}
where $z$ and $g^{\beta}_{ij}(z;\lbrace {\bf{ e}}\rbrace)$ are $E+i\delta$ and an auxiliary green function including spin-fluctuation, respectively. 
$\lambda^{\beta}_{i}(z; {\bf{ e}}_{i})$ and $\mu^{\beta}_{i} (z; {\bf{ e}}_{i} )$ are given as follows:
{
\begin{align}
\lambda^{\beta}_{i}(z; {\bf{ e}}_{i} )=({\Delta_{i}}({\bf{e}}_{i}))^{-1/2}(1+(\gamma_{i}({\bf e}_{i})-\beta){P}_{i}^{\gamma}(z;{\bf{e}}_{i}))({\Delta_{i}}({\bf{e}}_{i}))^{-1/2},
\end{align}
\begin{align}
\mu^{\beta}_{i}(z; {\bf{ e}}_{i} )=({\Delta_{i}}({\bf{e}}_{i}))^{-1/2}({P}_{i}^{\gamma}(z;{\bf{e}}_{i}))^{-1}P^{\beta}_{i}(z; {\bf e}_{i}),
\end{align}
\begin{align}
\bar \mu^{\beta}_{i}(z;  {\bf{ e}}_{i} )=P^{\beta}_{i}(z; {\bf e}_{i})({P}_{i}^{\gamma}(z;{\bf{e}}_{i}))^{-1}({\Delta_{i}}({\bf{e}}_{i}))^{-1/2},
\end{align}}
where 
 \begin{align}
{\Delta}_{i}^{-1/2}({\bf{e}}_{i})=U^{\dag}({\bf{e}}_{i})( {\Delta_{i}})^{-1/2}U({\bf{e}}_{i}),
 \end{align} 
 \begin{align}
({P}_{i}^{\gamma}(z;{\bf{e}}_{i}))^{-1}=U^{\dag}({\bf{e}}_{i})({P}_{i}^{\gamma}(z))^{-1}U({\bf{e}}_{i}),
 \end{align}  
  \begin{align}
{\gamma}_{i}({\bf{e}}_{i})=U^{\dag}({\bf{e}}_{i})( {\gamma_{i}}) U({\bf{e}}_{i}),
 \end{align} 
 \begin{align}
P_{i}^{\beta}(z;{\bf{e}}_{i})=U^{\dag}({\bf e}_{i})P^{\beta}_{i}(z) U({\bf e}_{i}),
\label{PFBRT}
\end{align}
\begin{align}
P^{\beta}_{i}(z)=P^{\gamma}_{i}(z)\lbrace1-[\beta-\gamma_{i}]P^{\gamma}_{i}(z)\rbrace^{-1},
\label{Trans}
\end{align}
\begin{align}
P^{\gamma}_{i}(z)=(\Delta_{i})^{-1/2}[z-C_{i}](\Delta_{i})^{-1/2}.
\label{PG}
\end{align}
Here, $\gamma_{i}$, $\Delta_{i}$, and $C_{i}$ are called potential parameters in the TB-LMTO method. 
 {The} $\beta$ values are summarized in several papers\cite{Sakuma2022,Turek,Kudrnovsky}.
 In this work, we neglect the SOC to calculate the $\Omega_{\textrm{eff}}(\lbrace{\bf{e}}\rbrace,T)$ and $\omega_{i}({\bf{e}}_{i},T)$.
 The effective grand potential {of electronic part} is expressed as
 \begin{align}
& \Omega_{\textrm{eff}}(\lbrace{\bf{e}}\rbrace,T)\sim \frac{1}{\pi} \int \textrm{d} \epsilon f(\epsilon,T,\mu) \int_{-\infty}^{\epsilon} \textrm{d} E \ \textrm{ImTr} \ G (z;\lbrace {\bf{ e}}\rbrace) \nonumber
 \\ &=-\frac{1}{\pi} \int \textrm{d} \epsilon f(\epsilon,T,\mu)\textrm{Im} \left[\textrm{Tr}\  \textrm{log}\  \lambda^{\beta}(\epsilon^{+};\lbrace {\bf{ e}}\rbrace) + \textrm{Tr} \ \textrm{log}\ g^{\beta}(\epsilon^{+};\lbrace {\bf{ e}}\rbrace) \right],
 \label{EFFGP}
 \end{align}
 where $f$ and $\mu$ { represent} the Fermi-Dirac function and the chemical potential, respectively. The trace is taken over with respect to sites $i$, orbitals $L$, and spin indices $\sigma$.
 From here, we expand $g^{\beta}(\epsilon^{+};\lbrace {\bf{ e}}\rbrace)$ with the auxiliary coherent Green function $\bar g^{\beta}(z)$, which is defined as follows:
 \begin{align}
\bar g^{\beta}(z)=\left(\bar P(z)-S^{\beta} \right)^{-1},
\label{CPAG}
\end{align}
where $S^{\beta}$ is given as 
\begin{align}
S^{\beta}=S(1-\beta S)^{-1}.
\end{align}
$S$ is a bare structure constant matrix\cite{Skriver}.  
The auxiliary  Green function $g^{\beta}(\epsilon^{+};\lbrace {\bf{ e}}\rbrace)$ is expanded as follows:
\begin{align}
g^{\beta}(z;\lbrace {\bf{ e}}\rbrace)=\bar g^{\beta}(z) \left(1+\Delta P (z;\lbrace {\bf{ e}}\rbrace)\bar g^{\beta} (z)\right)^{-1},
\label{expgf}
\end{align}
where
\begin{align}
\Delta P(z;\lbrace{\bf e}\rbrace)=P^{\beta}(z; \lbrace{\bf e}\rbrace)-\bar P(z),
\label{expandon}
\end{align}
and $\bar P$ is a coherent potential function. We also need to obtain $\bar P$ in a self-consistent manner (explained later).
Using Eq. \eqref{expgf}, Eq.\eqref{EFFGP} can be rewritten as follows:
\begin{align}
&\Omega_{\textrm{eff}}(\lbrace{\bf{e}}\rbrace,T)=-\frac{1}{\pi} \int \textrm{d} \epsilon \ f(\epsilon,T,\mu) \ \textrm{Im} \ \Biggl[\textrm{Tr}\  \textrm{log}\ \lambda^{\beta}(\epsilon^{+};\lbrace {\bf{ e}}\rbrace)\nonumber \\&+\textrm{Tr}\  \textrm{log} \ \bar g^{\beta}(z)-\textrm{Tr}\  \textrm{log} \left(1+\Delta P(z;\lbrace{\bf e}\rbrace)\bar g^{\beta}(z)\right)\Biggr].
\end{align}
By taking the trace with respect to site $i$, the grand potential can be expressed as follows\cite{Sakuma2022}:
\begin{align}
\Omega_{\textrm{eff}}(\lbrace{\bf{e}}\rbrace,T)=\Omega_{0}+\sum_{i} \Delta \Omega_{i} ({\bf e}_{i},T),
\end{align}
\begin{align}
\Delta \Omega_{i} ({\bf{e}}_{i},T)=\frac{1}{\pi} \textrm{Im} \int \textrm{d} E f(E,T,\mu) \textrm{Tr}_{L\sigma} \ \textrm{log} \left(1+\Delta P_{i}(z;{\bf e}_{i})\bar g^{\beta}_{ii}(z)\right).
\label{Domega}
\end{align}
Here, we used the fact that  ${\lbrace \bf e \rbrace}$ dependence of $\lambda^{\beta}(\epsilon^{+};\lbrace {\bf{ e}}\rbrace)$ vanishes in our case.
Therefore, $\omega_{i}({\bf{e}}_{i},T)$ can be expressed as follows:
\begin{align}
&\omega_{i}({\bf{e}}_{i},T) \nonumber \\&=\textrm{exp} \left(-\Delta\Omega_{i}({\bf{e}}_{i},T)/k_{\textrm{B}}T \right)/ \int \textrm{d} {\bf e}_{i} \textrm{exp} \left(-\Delta\Omega_{i} ({\bf{e'}}_{i},T)/k_{\textrm{B}}T\right),
\label{weight}
\end{align}
where $k_{\textrm{B}}$ is { the} Boltzmann constant. Finally, we need to determine { the} converged $\omega_{i}({\bf{e}}_{i},T)$ and $\bar P(z)$ self-consistently. 
The CPA condition to determine $\bar P(z)$ is given as:
\begin{align}
\int \textrm{d} {{\bf e}}_{i} \ \omega_{i}({\bf{e}}_{i},T)  \Delta P_{i}(z;{\bf e}_{i}) \left[1+ \Delta P_{i}(z;{\bf e}_{i})\bar g^{\beta}_{ii}(z)\right]^{-1} =0.
\label{CPAcon}
\end{align}
We use Eq. \eqref{CPAG}, Eq. \eqref{expandon}, Eq. \eqref{Domega}, Eq. \eqref{weight}, and Eq. \eqref{CPAcon} to obtain $\bar P_{i}$ and { the} converged $\omega_{i}({\bf{e}}_{i},T)$ in a self-consistent manner. \\
\ 
Once we obtain { the} converged $\omega_{i}({\bf{e}}_{i},T)$, we can calculate the SOP formula at finite temperatures as follows:
\begin{align}
&\delta E^{\textrm{2nd}} (T,{\bf n}) =\nonumber \\ &-\frac{1}{2\pi}\sum_{ij} \textrm{Im}\textrm{Tr}_{L\sigma}\int_{-\infty}^{{\infty}} \textrm{d} E \ f(E,\mu,T)\nonumber \\ &\times  \langle G_{ij}(z;\lbrace {\bf{ e}}\rbrace,{\bf n}) H^{\textrm{soc}}_{j}G_{ji}(z;\lbrace {\bf{ e}}\rbrace,{\bf n})H^{\textrm{soc}}_{i} \rangle_{\lbrace {\omega_{i}({\bf{e}}_{i},T)}\rbrace},
\label{FTSP}
\end{align}
where $H^{\textrm{soc}}$ and ${\bf n}$ { represent} the spin -orbit Hamiltonian and the magnetization direction, respectively. Similar expressions were used in several works\cite{Miura2022,Kobayashi2}.  
$\langle \cdots\rangle$ { denotes} the average {over $\lbrace{\bf e}\rbrace$} with {a weight of} $\omega_{i}({\bf{e}}_{i},T)$, { which is} given as follows:
\begin{align}
\langle \cdots \rangle_{\lbrace {\omega_{i}({\bf{e}}_{i},T)}\rbrace}= \prod_{i}  \int \textrm{d} {\bf e}_{i}\ \omega_{i}({\bf{e}}_{i},T) (\cdots).
\end{align}
 The rotation of { the} magnetization direction is expressed with the SO(3) rotation matrices $R({\bf n})$ as follows\cite{Yamashita2022,Ebert,Sakuma1}:
\begin{align}
S^{l_{1}l_{2}}_{m_{1}m_{2}}({\bf n})=\sum_{m_{3}m_{4}} R^{l_{1}*}_{m_{3}m_{1}}({\bf n}) S^{l_{1}l_{2}}_{m_{3}m_{4}} R^{l_{2}}_{m_{4}m_{2}}({\bf n}),
\label{rts}
\end{align}
where $*$ { denotes the} complex conjugate. We substitute Eq. \eqref{rts} into Eq. \eqref{CPAG} to express the rotation of the direction of magnetization.
\\ 
We can decompose Eq. \eqref{FTSP} into {onsite} $E_{ii}^{\textrm{2nd}}$ and {pair}  $E_{ij}^{\textrm{2nd}}$ contributions as follows:
\begin{align}
&E^{\textrm{2nd}}_{ii}(T,{\bf n})\nonumber \\& =-\frac{1}{2\pi} \textrm{Im}\textrm{Tr}_{L\sigma}\int  \textrm{d} { \bf e}_{i} \ \omega_{i}({\bf e}_{i},T)  \int_{-\infty}^{\infty} \textrm{d} E  f(E,T,\mu)  \nonumber \\ &\times \biggl\lbrace  H^{\textrm{soc}}_{i }\left(\lambda^{\beta}_{i}(z; {\bf{ e}}_{i} )+ \mu^{\beta}_{i} (z;  {\bf{ e}}_{i}) \bar g^{\beta}_{ii}(z,{\bf n}) \xi_{i}(z;{\bf{ e}}_{i},{{\bf n}})\bar \mu^{\beta}_{i} (z;  {\bf{ e}}_{i} )\right)\nonumber \\ \times  &\ H^{\textrm{soc}}_{i} \left(\lambda^{\beta}_{i}(z; {\bf{ e}}_{i})+ \mu^{\beta}_{i} (z; {\bf{ e}}_{i}) \bar g^{\beta}_{ii}(z,{\bf n}) \xi_{i}(z;{\bf{ e}}_{i},{{\bf n}}) \bar \mu^{\beta}_{i} (z;  {\bf{ e}}_{i})\right)  \biggr\rbrace,
\label{onsite1}
\end{align}
\begin{align}
&E^{\textrm{2nd}}_{ij}(T,{\bf n})\nonumber \\& =-\frac{1}{2\pi}\sum_{k} \textrm{Im}\textrm{Tr}_{L\sigma}\int  \textrm{d} { \bf e}_{i} \ \omega_{i}({\bf e}_{i},T) \int  \textrm{d} { \bf e'}_{j} \omega_{j}({\bf e'}_{j},T) \nonumber \\ &\times \int_{-\infty}^{\infty} \textrm{d} E  \ f(E,T,\mu)\ \lbrace \tilde H^{\textrm{soc}}_{i } \chi_{ik} (1-\Gamma \chi)^{-1}_{kj}\tilde H^{\textrm{soc}}_{j}  \rbrace.
\label{twosite12}
\end{align}
Here, we introduce $\xi_{i}(z;{\bf{ e}}_{i},{\bf {{\bf n}}})$, $\tilde \xi_{i}(z;{\bf{ e}}_{i},{{\bf n}})$, $\tilde H_{i}^{\textrm{soc}}$, $\Gamma$, and $\chi$, which are given as follows:
\begin{align}
\xi_{i}(z;{\bf{ e}}_{i},{{\bf n}})=\left[1+\Delta P_{i}(z;{\bf e}_{i})\bar g^{\beta}_{ii} (z,{{\bf n}})\right]^{-1},
\end{align}
\begin{align}
\tilde \xi_{i}(z; {\bf{ e}}_{i},{{\bf n}})=\left[1+\bar g^{\beta}_{ii} (z,{{\bf n}})\Delta P_{i}(z;{\bf e}_{i})\right]^{-1}.
\end{align}
\begin{align}
\tilde H^{\textrm{soc}}_{i}=\xi_{i}(z;{\bf{ e}}_{i},{{\bf n}}) \mu_{i}^{\beta} (z; {\bf{ e}}_{i}) H^{\textrm{soc}}_{i} \bar \mu_{i}^{\beta} (z;{\bf{ e}}_{i}) \tilde \xi_{i} (z; {\bf{ e}}_{i},{{\bf n}}),
\end{align}
\begin{align}
&\Gamma_{i}(z,T,{{\bf n}}) \nonumber \\&=\int \textrm{d} { \bf e}_{i}\  \omega_{i}({\bf e}_{i},T) \left[\Delta P^{\beta}_{i}(z; {\bf e}_{i})\tilde \xi_{i} (z; {\bf{ e}}_{i},{{\bf n}})\right] \left[\Delta P^{\beta}_{i}(z; {\bf e}_{i})\tilde \xi_{i}(z; {\bf e}_{i},{{\bf n}})\right],
\end{align}
\begin{align}
\chi_{ij}(z,{{\bf n}})=\bar g^{\beta}_{ij}(z,{{\bf n}})\bar g^{\beta}_{ji}(z,{{\bf n}}) (1-\delta_{ij}).
\end{align}
For evaluating {pair} contributions, we expand the Green function including the spin-fluctuation with the T-matrix to include the vertex correction terms.
Details regarding the derivation of the vertex correction terms are provided in several papers\cite{Butler,Carva,Hiramatsu2023}. \\
\ In practical calculations, we neglect the Fermi--Dirac distribution function in Eqs. \eqref{onsite1} and \eqref{twosite12}. { This} does not cause serious numerical errors.
 In { the present} study, the $K_{1}(T)$ part of the MAE at finite temperatures is defined as follows:
\begin{align}
&\epsilon_{\textrm{MAE}}(T)\nonumber \\&=\sum_{ij} \left( E^{\textrm{2nd}}_{ij}(T,\theta={\pi}/{2})-E^{\textrm{2nd}}_{ij}(T,\theta=0) \right)\sim K_{1}(T).
\end{align}
Using Eqs. \eqref{onsite1} and \eqref{twosite12}, we can investigate the site-resolved contributions in $K_{1}(T)$ and its temperature dependences. \\
%%%%%%%%%%%%%%%%%%%%%%%%%%%%%%%%%%%%%%%%%%%%%%%%%%%%%%%%%%%%%%%%%%%%%%%%%%
%%% 計算条件
%%%%%%%%%%%%%%%%%%%%%%%%%%%%%%%%%%%%%%%%%%%%%%%%%%%%%%%%%%%%%%%%%%%%%%%%%%
\ For calculation details, the lattice constants of each alloy are set to the same values used in a previous work\cite{Yamashita2022}. The number of $k$-points for each calculation is also the same as that in the previous work.
%%%%%%%%%%%%%%%%%%%%%%%%%%%%%%%%%%%%%%%%%%%%%%%%%%%%%%%%%%%%%%%%%%%%%%%%%%
%%%% Figure 1の説明
%%%%%%%%%%%%%%%%%%%%%%%%%%%%%%%%%%%%%%%%%%%%%%%%%%%%%%%%%%%%%%%%%%%%%%%%%%
\section{Results and discussions}
\begin{flushleft} 
\begin{figure}[htb]
\begin{center}
\includegraphics[clip,width=9.5cm]{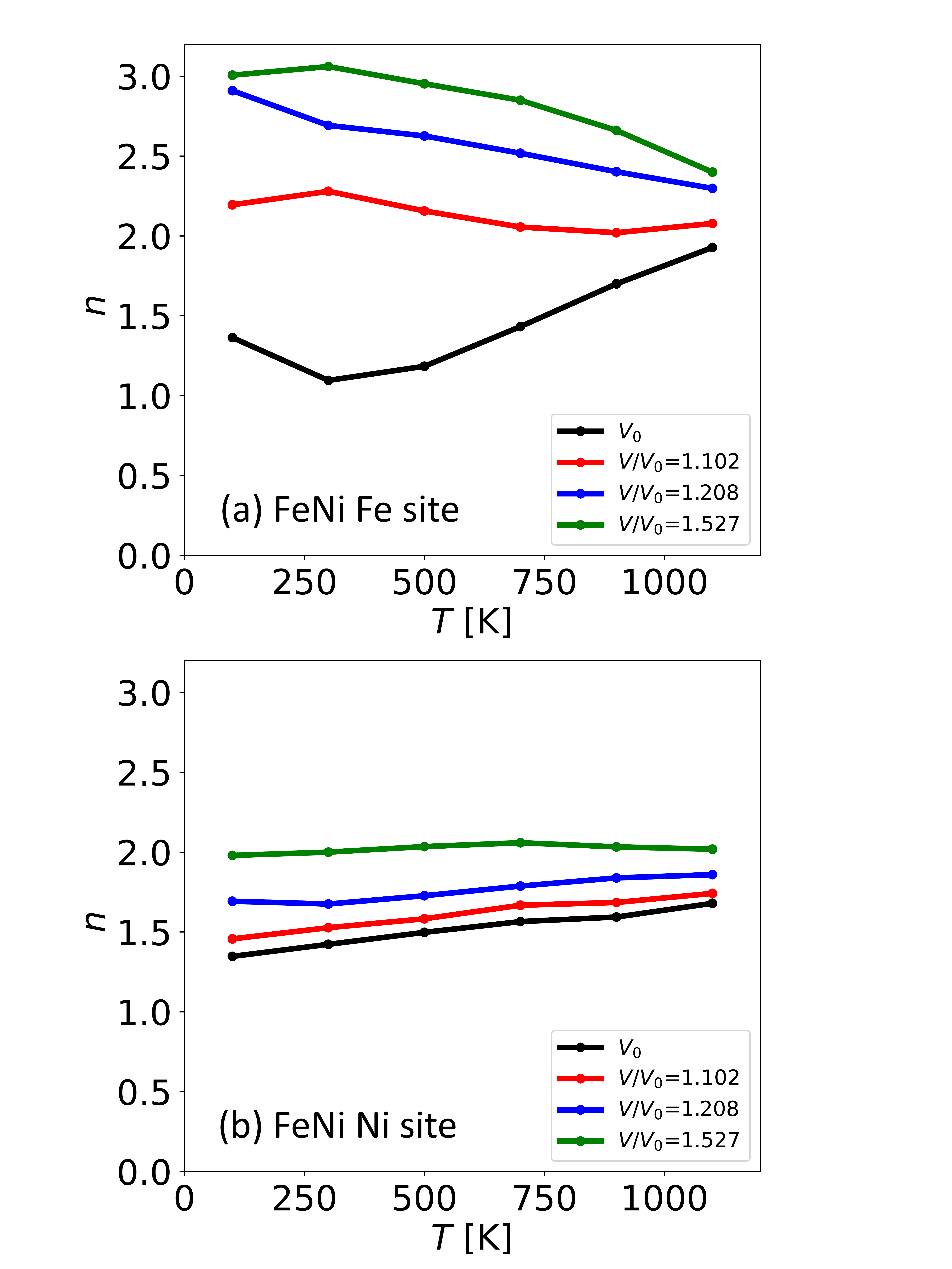}
\end{center}
\caption{Temperature dependences of the exponent $n$ of $K^{\textrm{on}}_{i}(T)/K^{\textrm{on}}_{i}(0)=\left(M_{i}(T)/M_{i}(0)\right)^n$ for Fe and Ni sites for various volumes $V$ of FeNi. Black line represents the results for the original volume. Other lines represent to the lattice-expanded results. $V_{0}$ is set to 22.6${\textrm{\AA}}^3$.}
\label{STTMAE4}
\end{figure}
\end{flushleft}

{{ To} examine the accuracy of the developed method, let us first { investigate} how the SOP method can reproduce the MAE obtained  { via} the force theorem (FT) in our previous work\cite{Yamashita2022}. Figure \ref{STTMAE3} shows the MAE calculated { using} the FT and the SOP method for FePt, MnAl, and FeNi. { Small differences between the results of the two methods are observed} for FePt and FeNi, whereas little difference { is observed for} in MnAl. For MnAl, the good agreement { is} reasonable, considering that the SOC of this system is { far weaker} than { those of} FePt and FeNi.
 { From}  this view-point,  { the} origin of the larger difference for FeNi  { compared with} FePt is not simple,  { because} the SOC in FeNi is  { weaker} than that in FePt.  { This may suggest} the peculiar $K_{1}(T)$ behavior  of FeNi, which will be discussed later.
  In total { summary}, the qualitative behaviors of the temperature dependence of the MAE in the previous work can well be reproduced by the SOP method focusing on $K_{1}(T)$. The difference between the FT and the SOP method { may arise} from the higher-order perturbation term, which contributes to $K_{2}(T)$. \\ 
\ The total onsite and pair contributions to $K_{1}(T)$ are also shown in Fig. \ref{STTMAE3}. For FePt, onsite and pair contribution have opposite { signs for the} whole temperature region. The onsite contribution is suppressed by the pair contribution, which leads to uniaxial anisotropy of $K_{1}(T)$. \\ 
\ { For} MnAl,  $K_{1}(T)$ is mostly dominated by the onsite contribution. The pair contribution makes a small correction to the $K_{1}(T)$. In this case, both contributions have a positive sign. \\
\ For FeNi, the onsite and pair contributions have similar { amplitudes, whereas the signs are opposite, as in} the case of FePt.
In addition, the peculiar behavior that $K_{1}(T)$ exhibits a plateau in the low-temperature region is found to be { due} to the cancellation of the variations of the total onsite and total pair terms with the temperature change. \\ 
\ {We stress here that even though the crystal structures are the same for these alloys, { the alloys differ with regard to the} breakdown of these SOP results into onsite and pair contributions. { In particular} for FePt and MnAl, despite the similar behavior of the temperature dependence of the total $K_{1}(T)$, the  onsite and pair { contributions differ significantly}. Furthermore, comparing FePt and FeNi { reveals} that the temperature dependences of onsite and pair contributions are { differ significantly} between these two alloys. These characteristics of the $K_{1}(T)$ of each alloy can be recognized via the present SOP theory at finite temperatures, which we believe is one of the advantage of this approach. 
\\ \ Here, to investigate the characteristics of the temperature dependence of $K_{1}(T)$ in the itinerant electron magnets, we compare those results with the results expected from the spin model.
The Hamiltonian is given by the XXZ model\cite{Yamashita2022,Mryasov,Evans,Cuadrado} as follows:
\begin{align}
H=-\sum_{( i,j ) } 2 J_{ij} \vec S_{i}\cdot \vec S_{j}-\sum_{( i,j ) } D_{ij}S^{z}_{i} S^{z}_{j}-\sum_{i} D_{i}{(S^{z}_{i} )}^2,
\label{XXZ}
\end{align}
where $\vec S_{i}$, $J_{ij}$, $D_{i}$, and $D_{ij}$ are a classical spin vector, an exchange coupling constant, a single-site anisotropy coefficient, and  a two-site anisotropy coefficient, respectively.  
As presented in our previous work\cite{Yamashita2022}, analysis with this model requires $D_{i}>0$ and $\sum_{j}D_{ij}>0$ for FePt and MnAl and $D_{i}<0$ and $\sum_{j}D_{ij}>0$ for FeNi to reproduce the temperature dependence of the total MAE. However, if one naively assumes that the $D_{i}$ and $D_{ij}$ terms correspond to onsite and pair contributions, respectively, at first glance, the signs of the terms in the XXZ model used in the previous work are not consistent with the results in Fig. \ref{STTMAE3}, except for the case of MnAl. This mismatch originates from the fact that the temperature dependences of the MAE from the $D_{i}$ and $D_{ij}$ terms in the XXZ model behave as approximately $\propto M^3(T)$\cite{Callen1,Callen2} and $\propto M^2(T)$\cite{Mryasov}, respectively, regardless of the signs and amplitudes of the parameters $D_{i}$ and $D_{ij}$, whereas those of the onsite and pair terms in the SOP method do not necessarily follow such simple rules but exhibit various behaviors depending on the system. For this reason, to reproduce the peculiar behavior of the total MAE of FeNi, the XXZ model has no other choice than to set $D_{i}<0$ and $\sum_{j} D_{ij}>0$; however, the onsite and pair contributions can produce such behavior with opposite signs from the XXZ model. %and different temperature dependence from the XXZ model. 
 Thus, the results in Fig. \ref{STTMAE3} imply that the spin model is too simple and is insufficient to express the temperature dependence of the MAE of itinerant magnets.}
\begin{flushleft} 
\begin{figure}[H]
\begin{center}
\includegraphics[clip,width=9.5cm]{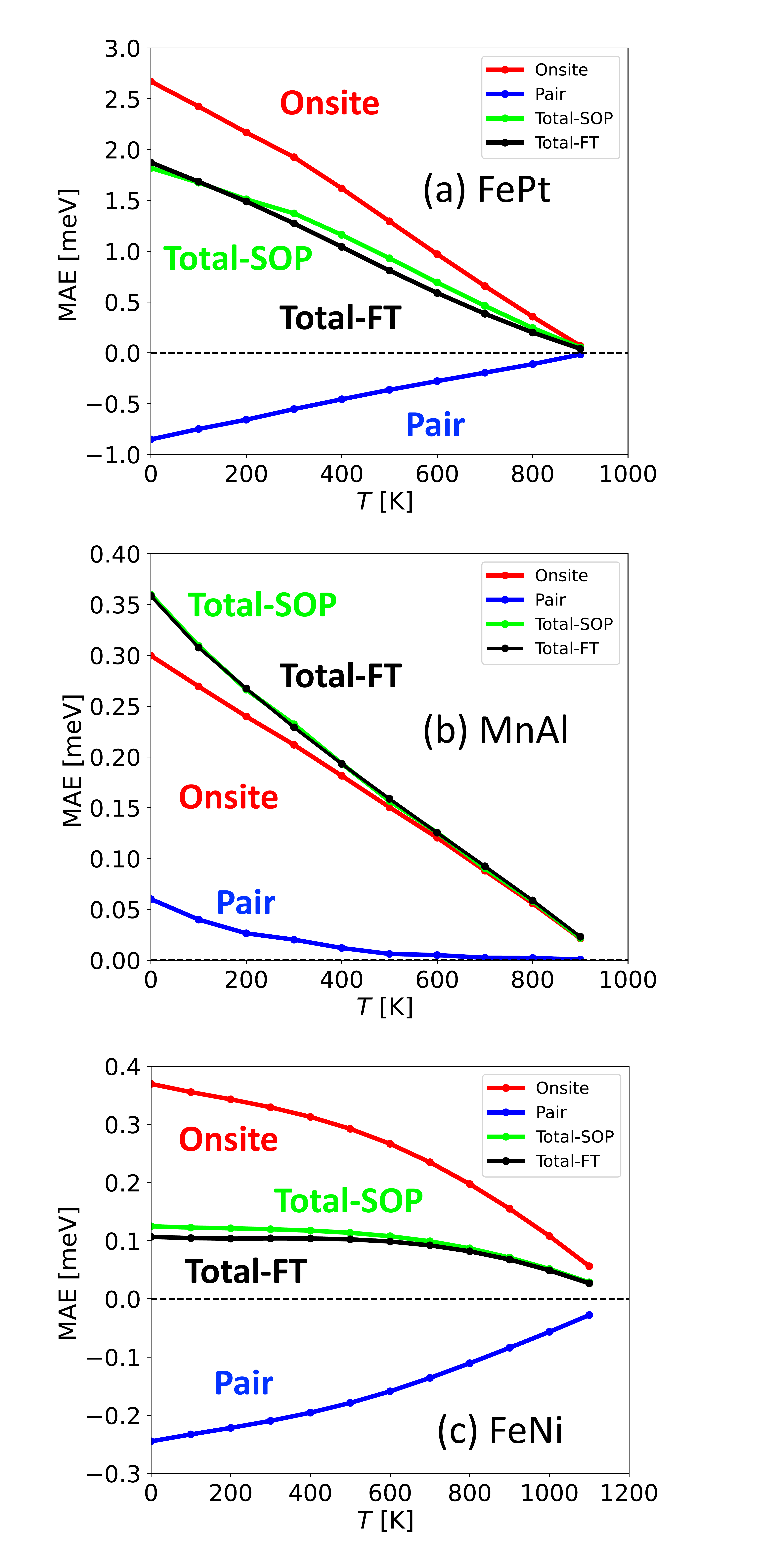}
\end{center}
\caption{Temperature dependence of the total $K_{1}(T)$ calculated { via} the developed finite-temperature SOP method including vertex correction terms for (a) FePt, (b) MnAl, and (c) FeNi, { which are} indicated by green lines. For comparisons, the results calculated { using the FT reported by} Yamashita et al.\cite{Yamashita2022} are shown { as} black lines. In addition, the total onsite and total pair contributions are { shown as} red and blue lines, respectively.}
\label{STTMAE3}
\end{figure}
\end{flushleft}
%%%%%%%%%%%%%%%%%%%%%%%%%%%%%%%%%%%%%%%%%%%%%%%%%%%%%%%%%%%%%%%%%%%%%%%%%%
%%%% Figure 2の説明
%%%%%%%%%%%%%%%%%%%%%%%%%%%%%%%%%%%%%%%%%%%%%%%%%%%%%%%%%%%%%%%%%%%%%%%%%%
 \ { T}o examine each contribution in detail, we decompose the SOP results into each onsite and pair contribution.  The breakdown of these contributions is shown in Fig. \ref{STTMAE2} for all the alloys. For FePt, the results are shown in Fig. \ref{STTMAE2} (a). The total onsite contribution in FePt is mostly { from} of Pt, and the Fe contribution is far smaller. In addition, it is found that the negative contribution of the total pair term in Fig. \ref{STTMAE3} (a) mainly comes from the Fe-Pt pair, and it is suppressed by the positive contribution from Pt-Pt pairs. This leads to a negative total pair contribution. {In this alloy, the onsite and pair contributions related to Pt { significantly affect} the temperature dependence of $K_{1}(T)$. } \\
 \ The results for MnAl in Fig. \ref{STTMAE2} (b) indicate that all the contributions are positive and that the situation regarding the total onsite term is similar to that for FePt. It is mostly dominated by the Mn onsite contribution. The second-largest contribution is the Mn-Mn positive pair contribution{, and the other contributions} are negligible. {Thus, the onsite and pair contributions of Mn determine the temperature dependence of  $K_{1}(T)$ in MnAl.} \\
 \ For FeNi, as shown in fig. \ref{STTMAE2} (c), the onsite contributions from Fe and Ni are positive and have similar amplitudes, leading to a total positive onsite contribution. We  also find that the negative contribution of the total pair term mostly comes from a pair of different atoms, i.e., Fe-Ni. While the Fe-Ni contribution is suppressed by other positive pair contributions, it finally leads to negative finite total pair contributions.} \\ 
 \ {From these results, although most of the pair contributions are suppressed by other pairs and the net contribution becomes small, the pair contributions play important roles { over the}  whole temperature region, particularly for FePt and FeNi. The importance of the pair contributions was also { investigated} { by} Ke\cite{Ke2} with the SOP method at 0 K. In this work, we confirmed that the pair contributions to $K_{1}(T)$ { significantly affect} the temperature dependence of $K_{1}(T)$ at not only 0 K but also finite temperatures.}
 \begin{flushleft} 
\begin{figure}[htb]
\begin{center}
\includegraphics[clip,width=9.5cm]{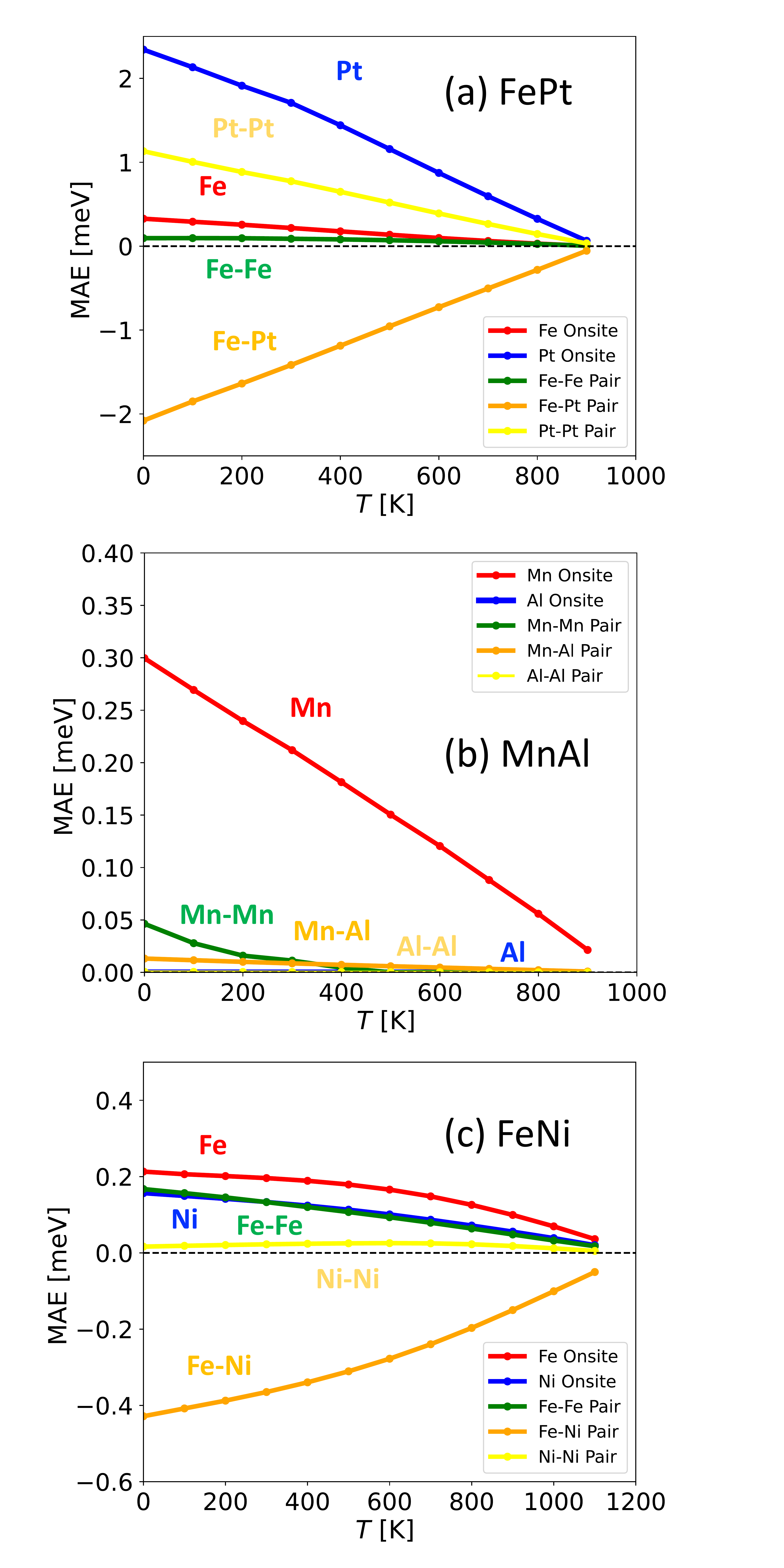}
\end{center}
\caption{Temperature dependence of the site-resolved contributions to $K_{1}(T)$ for (a) FePt, (b) MnAl, and (c) FeNi.}
\label{STTMAE2}
\end{figure}
\end{flushleft}
%%%%%%%%%%%%%%%%%%%%%%%%%%%%%%%%%%%%%%%%%%%%%%%%%%%%%%%%%%%%%%%%%%%%%%%%%%
%%%% Figure 3の説明
%%%%%%%%%%%%%%%%%%%%%%%%%%%%%%%%%%%%%%%%%%%%%%%%%%%%%%%%%%%%%%%%%%%%%%%%%%
{\ Finally, we virtually expand the lattice of FeNi to investigate the influence of electron itineracy on the onsite contribution $K^{\textrm{on}}_{i}(T)$. Here, we fit $K^{\textrm{on}}_{i}(T)$ by assuming the relation $K^{\textrm{on}}_{i}(T)\propto M_{i}(T)^n$ and investigate the temperature dependence of the exponent $n$ at each site.  
As mentioned previously, the onsite contributions $K^{\textrm{on}}_{i}(T)$ should follow the relation $K^{\textrm{on}}_{i}(T)\propto M_{i}(T)^3$ if the localized spin model is suitable to explain the temperature dependence of $K^{\textrm{on}}_{i}(T)$. We briefly examine the validity of the single-site anisotropy term, which is the simplest term, for the temperature dependence of the MAE in the itinerant magnets.
In Fig. \ref{STTMAE4}, the temperature dependences of $n$ of the onsite contributions for FeNi with various volumes are shown. If we expand the lattice, the $n$ value of $K^{\textrm{on}}_{\textrm{Fe}}(T) \propto M_{\textrm{Fe}}(T)^n$ { increases, and it reaches 3 in the low-temperature} region. However, the $n$ value of $K^{\textrm{on}}_{\textrm{Ni}}(T) \propto M_{\textrm{Ni}}(T)^n$ is always far from 3 and is not changed drastically. If we use a spin model with the single-site anisotropy term to explain the temperature dependence of the onsite contributions, the $n$ value must be fixed to 3 in the low-temperature region regardless of the sign and amplitude of $D_{i}$. 
In addition, if the lattice is expanded, the $D_{i}$ and $D_{ij}$ values are expected to change, and these temperature dependences are not changed in the localized spin model.
However, our results imply that not only changing the value of $D_{i}$ but also changing the temperature dependence of the onsite term itself with expanding the lattice. Therefore, we can again conclude that if we assume Eq. \eqref{XXZ} to explain the temperature dependence of the MAE, even the single-site anisotropy term in Eq. \eqref{XXZ} may not always be sufficient to describe the temperature dependence of $K^{\textrm{on}}_{i}(T)$ in the itinerant ferromagnets.} 
 %%%%%%%%%%%%%%%%%%%%%%%%%%%%%%%%%%%%%%%%%%%%%%%%%%%%%%%%%%%%
 %Summary
 %%%%%%%%%%%%%%%%%%%%%%%%%%%%%%%%%%%%%%%%%%%%%%%%%%%%%%%%%%%%
 \section{Summary}
 In summary, we developed a finite-temperature SOP method to describe the temperature dependence of $K_{1}(T)$ and applied it to $L1_{0}$-FePt, MnAl, and FeNi. We confirmed that the developed method can reproduce the results of a previous work\cite{Yamashita2022}. We also investigated the onsite and pair contributions to $K_{1}(T)$ with the developed method. We showed that not only the onsite contributions but also the pair-contributions { significantly affect} the temperature dependence of  $K_{1}(T)$. In particular, the unique behavior of $K_{1}(T)$ for FeNi is attributed to the competition of onsite and pair-site contributions. In addition, for some results, it is found that the signs of onsite and pair contributions do not agree with the conditions used in the previous work\cite{Yamashita2022}. {Finally, we investigated the effect of electron itineracy for the temperature dependence of the $K^{\textrm{on}}_{i}(T)$ of FeNi while expanding the lattice parameters. We found that the exponent $n$ of the onsite contributions of both atoms depends on the volume and is not fixed to 3, which is expected from a spin model. From the above, our results imply that the XXZ model, even the single-site anisotropy term in this model, is insufficient for the itinerant ferromagnets.} 
\begin{acknowledgments}
S.Y. acknowledges Dr. Ryoya Hiramatsu, Dr. Yusuke Masaki, and Prof. Dr. Hiroaki Matsueda of Tohoku University for fruitful discussions and support from GP-Spin at Tohoku University, Japan.
S.Y. also appreciates Dr. Juba Bouaziz and Prof. Dr. Stefan Bl\"ugel of Forschungszentrum J\"ulich for constructive comments on this work. 
This work was supported by JSPS KAKENHI Grant Number  JP19H05612 in Japan.
\end{acknowledgments}

\end{document}